\documentclass[conference]{IEEEtran}
\IEEEoverridecommandlockouts
\usepackage{cite}

\usepackage{newtxtext}
\usepackage[varg]{newtxmath}
\usepackage{listings}
\ifCLASSOPTIONcompsoc
    \usepackage[caption=false, font=normalsize, labelfont=sf, textfont=sf]{subfig}
\else
\usepackage[caption=false, font=footnotesize]{subfig}
\fi
\usepackage[hyphens]{url}
\usepackage[hidelinks]{hyperref}
\hypersetup{breaklinks=true}
\usepackage{algorithm}
\usepackage{algpseudocode}
\usepackage{soul}
\usepackage{cite}

\makeatletter
\newcommand*{\algrule}[1][\algorithmicindent]{\makebox[#1][l]{\hspace*{.5em}\vrule height .75\baselineskip depth .25\baselineskip}}%

\newcount\ALG@printindent@tempcnta
\def\ALG@printindent{%
    \ifnum \theALG@nested>0
        \ifx\ALG@text\ALG@x@notext
            \addvspace{-3pt}
        \else
            \unskip
            \ALG@printindent@tempcnta=1
            \loop
                \algrule[\csname ALG@ind@\the\ALG@printindent@tempcnta\endcsname]%
                \advance \ALG@printindent@tempcnta 1
            \ifnum \ALG@printindent@tempcnta<\numexpr\theALG@nested+1\relax
            \repeat
        \fi
    \fi
    }%
\usepackage{etoolbox}
\patchcmd{\ALG@doentity}{\noindent\hskip\ALG@tlm}{\ALG@printindent}{}{\errmessage{failed to patch}}
\makeatother

\algrenewcommand\algorithmicrequire{\textbf{Input:}}
\algrenewcommand\algorithmicensure{\textbf{Output:}}
\usepackage{soul}

\usepackage{textcomp}
\renewcommand\IEEEkeywordsname{Keywords}
\usepackage[pdftex]{graphicx,xcolor}
\usepackage{epstopdf}
 \DeclareGraphicsExtensions{.png,.pdf}
\usepackage[caption=false]{subfig}
\usepackage{float}
\usepackage{listings}
\usepackage{url}

\definecolor{mGreen}{rgb}{0,0.6,0}
\definecolor{mGray}{rgb}{0.5,0.5,0.5}
\definecolor{mPurple}{rgb}{0.58,0,0.82}
\definecolor{mNavy}{rgb}{0,0.5,1.0}
\definecolor{cRoutine}{rgb}{1.5, 0.3, 0.0}
\definecolor{backgroundColour}{rgb}{0.95,0.95,0.92}
\lstdefinestyle{CStyle}{
    language=C,
    basicstyle=\ttfamily\tiny,
    frame = single, 
    commentstyle=\color{mGreen},
    keywordstyle=\color{blue},
    numberstyle=\tiny\color{mGray},
    stringstyle=\color{mPurple},
    breakatwhitespace=false,         
    breaklines=true,                 
    captionpos=b,                    
    keepspaces=true,                 
    numbers=left,                    
    numbersep=5pt,                  
    showspaces=false,                
    showstringspaces=false,
    showtabs=false,                  
    tabsize=2
}
\floatstyle{plaintop}
\restylefloat{table}
\def\BibTeX{{\rm B\kern-.05em{\sc i\kern-.025em b}\kern-.08em
    T\kern-.1667em\lower.7ex\hbox{E}\kern-.125emX}}

\usepackage{geometry}
\makeatletter
\newcommand{\newlineauthors}{%
  \end{@IEEEauthorhalign}\hfill\mbox{}\par
  \mbox{}\hfill\begin{@IEEEauthorhalign}
}
\makeatother

\begin{document}
\bstctlcite{IEEEexample:BSTcontrol}

\title{A Reorder Trick for Decision Diagram Based Quantum Circuit Simulation\\

}

\author{\IEEEauthorblockN{1\textsuperscript{st} Jingcheng Shen}
\IEEEauthorblockA{\textit{School of Computer Science and Technology} \\
\textit{Chongqing University of Posts and Telecommunications}\\
Chongqing, China \\
shenjc@cqupt.edu.cn}
\and
\IEEEauthorblockN{2\textsuperscript{nd} Linbo Long}
\IEEEauthorblockA{\textit{School of Computer Science and Technology} \\
\textit{Chongqing University of Posts and Telecommunications}\\
Chongqing, China \\
longlb@cqupt.edu.cn}
\and
\IEEEauthorblockN{3\textsuperscript{rd} Masao Okita}
\IEEEauthorblockA{\textit{Graduate School of Information Science and Technology} \\
\textit{Osaka University}\\
Osaka, Japan \\
okita@ist.osaka-u.ac.jp}
\and
\IEEEauthorblockN{4\textsuperscript{th} Fumihiko Ino}
\IEEEauthorblockA{\textit{Graduate School of Information Science and Technology} \\
\textit{Osaka University}\\
Osaka, Japan \\
ino@ist.osaka-u.ac.jp}
}

\maketitle
\maketitle

\begin{abstract}
Quantum computing is a hotspot technology for its potential to accelerate specific applications by exploiting quantum parallelism. However, current physical quantum computers are limited to a relatively small scale, simulators based on conventional machines are significantly relied on to perform quantum computing research. The straightforward array-based simulators require a tremendous amount of memory that increases exponentially with respect to the number of qubits. 
To mitigate such computing resource concerns,  decision diagram based simulators were proposed that can efficiently exploit data redundancies in quantum states and operations.
In this paper, we study two classes of quantum circuits on which the state-of-the-art decision diagram based simulators failed to perform well in terms of simulation time. We also propose a simple and powerful reorder trick to boost the simulation of such quantum circuits. Preliminary evaluation results demonstrate the usefulness of the proposed trick. Especially, for the Quantum Phase Estimation circuits, the proposed trick achieved speedups up to 313.6$\times$ compared to a state-of-the-art approach that relies on an auxiliary tool to optimize simulation order.  
\end{abstract}


\section{Introduction}

Quantum computing is emerging as a research hotspot because of its potential for massively parallel processing. 
Quantum algorithms exploit quantum mechanisms such as superposition and entanglement of quantum bits (i.e. qubits) to outperform conventional machines in solving certain tasks. 
For instance, Shor's algorithm \cite{shor} is considered efficient in terms of complexity in factoring integers.   
Nonetheless, current realizations of quantum computers such as \cite{ibm,google} suffer from physical limits such as a relatively small number of qubits and a low fidelity due to noises, conventional machine based simulators are thus widely used in quantum computing for purposes such as designing quantum circuits (QCs) and validating quantum algorithms.

Array-based simulators such as \cite{qhipster,qulacs,quest} are a straightforward solution, performing the simulation of a QC as a sequence of linear algebra calculations. That is, the quantum state is represented as a vector (i.e., array) and operations such as quantum gates are represented as matrices. Updating the quantum state is done by performing a matrix-vector multiplication with the operation matrix and the state vector. However, the memory consumption increases exponentially with respect to the number of qubits, and the array-based QC simulation is thus limited to several tens of qubits even with supercomputers \cite{wu2021accelerating}.

To alleviate the concern of a tremendous amount of memory use, decision diagram (DD) based simulators have been proposed \cite{niemann2015qmdds,zulehner2018advanced}. Such a DD-based simulator allows a compact representation of the quantum states and operations by reducing data redundancies in the state vectors and operation matrices. 
Many successful cases demonstrate that the DD-based simulators make it possible to simulate circuits of a large number (in some cases even more than 100) of qubits using a machine with moderate specifications.

In this paper, we studied two important classes of QCs, i.e., the entangled Quantum Fourier Transform (QFT) and Quantum Phase Estimation (QPE) circuits, on which the DD-based simulators failed to perform well in terms of simulation time. We propose a simple and effective trick for the DD-based simulator to fast simulate such circuits.

The reminder of the paper is organized as follows. Section~\ref{sec:rel} summarizes the related work on DD-based QC simulation.
Section~\ref{sec:bgd} gives a minimal knowledge of DD-based QC simulation. 
Section~\ref{sec:prop} shows the details of the proposed method. 
Experimental results are provided in Section~\ref{sec:exp}. Finally, we conclude the paper in Section~\ref{sec:conc}.   

\section{Related Work}
\label{sec:rel}

Niemann \textit{et al.}\cite{niemann2015qmdds} laid the very foundation, QMDDs, the first practical DD-based QC simulator. In their work, they realized a compact and canonic representation of quantum functionality. The QMDDs simulator provides the basis for sophisticated quantum computing applications such as solutions for synthesis, simulation, and verification.

Zulehner \textit{et al.}\cite{zulehner2018advanced} highlighted the up-to-date version of \cite{niemann2015qmdds} by showcasing the advancement of the DD-based simulator compared to other state-of-the-art array- and graph-based simulators. They revisited the basics of quantum computation and developed a simulation approach that exploits redundancies in the respective quantum state and operation descriptions. 

Burgholzer \textit{et al.}\cite{cotengra} studied the importance of choosing a good simulation path when using DD-based simulators. They borrowed contraction order strategies from
the domain of tensor networks to optimize the order of simulating a QC with DD-based simulators, showing a significant speedup compared to the straightforward fashion used by previous DD-based simulations. Their approach involves the use of an auxiliary library to optimize the simulation order before simulating a QC, whereas our proposed trick is simpler in use. Moreover, our trick outperforms their work in simulating the QPE circuits. 

Burgholzer \textit{et al.}\cite{hybrid} proposed a hybrid Schrödinger-Feynman technique to parallelize DD-based simulation. Such a technique makes a QC branch into doppelgängers by breaking the controlled quantum gates. The doppelgänger circuits can be simulated simultaneously and the results will be summed up to get the entire picture. However, the number of doppelgänger circuits increases exponentially with respect to the number of controlled gates we must break. Therefore, this technique can only be applied to very shallow circuits.

\section{Background}
\label{sec:bgd}
\begin{figure*}[tb!]
    \centering
    \includegraphics{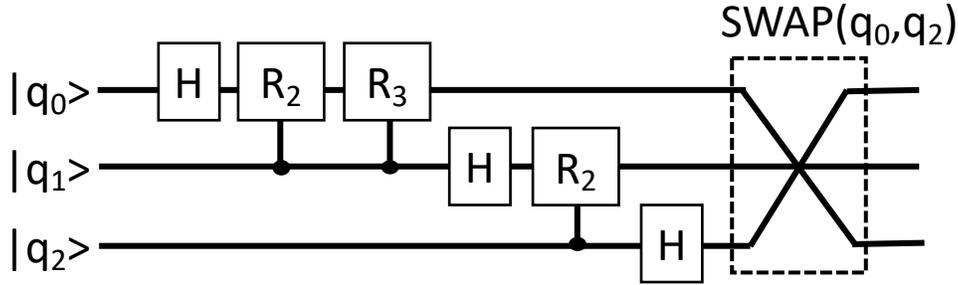}
    \caption{A three-qubit QFT circuit. Note that a swap gate exists at the end. Such final swap gates limit the performance of entangled QFT circuits.}
    \label{fig:qft}
\end{figure*}

To make this article self-contained, this section gives a basic knowledge of DD-based QC simulation.  
\subsection{Quantum Circuit Simulation}
\label{sec:qc}
The state of a single-qubit system can be given as a vector $[\alpha,\beta]^{T} \equiv \alpha [1,0]^{T} + \beta [0,1]^{T} \equiv \alpha$$|$0$>$$+$$\beta$$|$1$>$, where $\alpha^{2}$ is the probability to get the basic state $|$0$>$ and $\beta^{2}$, the probability to get $|$1$>$. Note that $\alpha,\beta \in \mathbb{C}$ and $\alpha^{2}$+$\beta^{2}$=1.

For an $N$-qubit system, the state $|$$\phi$$>$ is the tensor product ($\otimes$) of $N$ single-qubit states, i.e., $|$$\phi$$>$=$\otimes_{i=0}^{N-1}[\alpha_{i},\beta_{j}]^{T}$, where $\alpha_{i},\beta_{i} \in \mathbb{C}$. Note that the sum of squares of all the entries (a.k.a. probability amplitudes) in $|$$\phi$$>$ is equal to 1 to make a probabilistic sense.

Operations in QC simulation can be represented as matrices. 
Therefore, applications of operations to qubits can be handled as calculations in Linear Algebra. 
For instance, the Hadamard gate $H$, one of the most important single-qubit gates, can be described as $\frac{1}{\sqrt{2}}
\begin{bmatrix}
1 & 1 \\
1 & -1 \\
\end{bmatrix}$
. Applying $H$ to $|$0$>$ is a simple matrix-vector multiplication: $H|$0$>\equiv\frac{1}{\sqrt{2}}[1,1]^{T}\equiv\frac{1}{\sqrt{2}}|$0$>$+$\frac{1}{\sqrt{2}}|$1$>$.

Array-based approaches are straightforward for QC simulation. However, the major drawback of such approaches is the memory consumption which grows exponentially with the increasing number of qubits. Actually, during simulation, the array that represents the state vector has many repeated data patterns that can be exploited by techniques such as data compression \cite{wu2019full,wu2021accelerating,zhao2022q}. Moreover, the DD-based simulator \cite{niemann2015qmdds,zulehner2018advanced} can inherently make use of such data compressibility.

\subsection{Decision Diagram Based Simulation}
\label{sec:ddsim}
As the name implies, the DD-based simulator \cite{niemann2015qmdds,zulehner2018advanced} maintains the state vector as DDs. In general, qubits are represented as nodes whereas the probability coefficients are dynamically computed and assigned to the edges between nodes. The probability amplitude for a specific state (or sub-state) can be obtained by taking the product of all the coefficients along the according path from the root node to the target node. Note that states (or sub-states) with the same probability amplitudes can be merged into the same sub-diagrams, thus reducing the data redundancy. 

\section{Methodology}
\label{sec:prop}
\begin{table*}[tb!]
    \centering
       \caption{Experimental Evaluations}
    \begin{tabular}{|l|l|l|l|l|}
    \hline
        Benchmark &  No. of Qubits  & Baseline & DDSIM w/ CoTenGra & DDSIM w/ reorder \\\hline
         Entangled QFT & 17  &  42.8 & 0.6 & \textbf{0.6}\\ 
                       & 18  &  395.6 &\textbf{0.9} & 1.9\\ 
                       & 19  & 5836.2 & \textbf{3.7}& 7.2\\ \hline
         QPE           & 17  & 27.2& 7.1 &  \textbf{0.2}\\ 
                       & 18  & 118.1 &14.2 & \textbf{0.2}\\ 
                       & 19  & 1837.0&282.2 & \textbf{0.9}\\ \hline
    \end{tabular}
    \label{tab:eval}
\end{table*} 
Although the DD-based simulator can significantly reduce the memory footprint of QC simulation, we notice that the simulator failed to perform well in terms of execution time on two relevant classes of circuits, namely, the entangled QFT and the QPE circuits. We surmise that the long execution time was caused by maintaining complex DD structures. In this section, we describe an optimization trick that are both simple and efficient for such circuits.  
\subsection{Eliminating Swap Gates for Entangled QFT}
\label{sec:elim}
 In studying the entangled QFT circuits, we used a ``trial-and-error'' heuristic to pinpoint the bottleneck of simulating such circuits. Precisely, we deleted gates from a circuit repeatedly to observe the change of simulation time. Finally, we observed that the sequence swap gates in the rear of the entangled QFT circuits consumed most of the simulation time. 
 Such gates swap each pair of qubits that are symmetric in the circuit. For instance, for a circuit of $n$ qubits, the qubit 0 will be swapped with the qubit $n-1$ at the end of simulation.
 Figure~\ref{fig:qft} illustrates a simple QFT circuit for better understanding.
 
 However, the index of a qubit is just used for users to refer to and has no special meaning.
 That is, if we know which qubit we want to refer to, the index can be arbitrary and the swap gates are thus unnecessary. Therefore, we can eliminate the sequence of swap gates in the rear of a QFT circuit. For example, if we delete the swap gate applied to the qubit 0 and the qubit $n-1$, we effectively reorder the two qubits. We must be careful when referring to the reordered qubits. If we want to query the amplitude of the state $|q_{0}xxxxxq_{n-1}$$>$, we must query $|q_{n-1}xxxxxq_{0}$$>$ instead. Note that the qubit reorder technique is also used in array-based simulation to improve the data locality for matrix-vector multiplications \cite{zhang2021hyquas}.    
 
\subsection{Qubit Reorder for QPE}
\label{sec:qreorder}

The QPE is another class of circuits on which the DD-based simulator does not perform well. Therefore, we want to take advantage of the observation of Section~\ref{sec:elim} because the QPE circuits are relevant to the QFT. Nevertheless, the QPE circuits utilize a variant of QFT, namely, inverse QFT, which scatters the swap gates throughout the circuit instead of putting them in the end of the circuit. Subsequently, we cannot simply delete such swap gates from the circuit because other gates may depend on them.

Therefore, we proposed an explicit reorder trick. If we cannot directly delete the swap gate applied to the qubit $i$ and the qubit $j$, we can instead replace the qubit $i$ with the qubit $j$ and vice versa. That is, all the gate take the form of $GATE(...,i,...,j,...)$ will be changed to $GATE(...,j,...,i,...)$. In doing so, we can effectively delete swap gates from the QPE circuits.

\section{Experimental Results}
\label{sec:exp}
We compare our reorder trick with the baseline \cite{zulehner2018advanced} and an state-of-the-art approach \cite{cotengra} that utilizes an auxiliary tool, the CoTenGra \cite{gray2021hyper} library, to optimize the simulating order before simulating a QC.
In the experiments, we used the entangled QFT and QPE circuits provided by the MQT Bench \cite{mqtbench}. A MacBook Pro laptop with an Apple M1 CPU was used to perform the experiments. The operation system is macOS Monterey 12.0.1 and the version of MQT DDSIM simulator \cite{zulehner2018advanced} is 1.12.3.  

As shown in Table \ref{tab:eval}, our reorder trick helped DDSIM fast simulate entangled QFT and QPE circuits. As for entangled QFT circuits, although DDSIM with qubit reorder was slower than a previous work \cite{cotengra}, it is simpler to use, avoiding the use of any auxiliary tools such as the CoTenGra library. As for QPE circuits, the reorder trick can even beat the CoTenGra-based solution, achieving speedups up to 313.6$\times$. 

\section{Conclusions}
\label{sec:conc}
 In this work, we proposed a qubit reorder trick for DD-based simulation, especially to boost the simulation of QFT-related circuits on which the current DD-based simulator does not perform very well. We studied the effects of directly deleting final swap gates for the entangled QFT circuits and explicitly reordering qubits for the QPE circuits. 
 Experimental evaluations show that the qubit reorder trick is both simple and effective. Especially, the qubit reorder trick achieved speedups up to 313.6$\times$, compared to a state-of-the-art approach based on an auxiliary optimization tool \cite{cotengra}.
\section*{Acknowledgment}
This study was supported in part by the Japan Society for the
Promotion of Science KAKENHI under grants 20K21794.
\bibliographystyle{IEEEtran} 
\bibliography{cite}

\end{document}